\documentclass[final,5p,times,twocolumn,bibtex,natbib]{elsarticle}
\RequirePackage[T1]{fontenc}
\RequirePackage{mathptmx}
\usepackage{amsmath, amssymb}
\usepackage{slashed}
\usepackage{graphicx}
\usepackage{amsmath}
\usepackage{amssymb}
\usepackage{bbold}
\usepackage{wrapfig}
\usepackage{color}
\usepackage{caption}
\usepackage{subcaption}
\usepackage{hyperref}
\usepackage{multirow}
    \usepackage[vertfit]{breakurl} 
\usepackage{fullpage}
\journal{}
\DeclareMathOperator{\Tr}{Tr}
\begin{document}
\begin{frontmatter}
\title{Magnetic Vortex Lattices in Finite Isospin Chiral Perturbation Theory}
\author{Prabal Adhikari}
\ead{adhika1@stolaf.edu}
\date{\today}
\address{Physics Department, Faculty of Natural Sciences and Mathematics, St. Olaf College, 1520 St. Olaf Avenue, Northfield, MN 55057, United States}
\begin{abstract}
We study finite isospin chiral perturbation theory ($\chi$PT) in a uniform external magnetic field and find the condensation energy of magnetic vortex lattices using the method of successive approximations (originally used by Abrikosov) near the upper critical point beyond which the system is in the normal vacuum phase. The difference between standard Ginzburg-Landau (GL) theory (or equivalently the Abelian Higgs model) and $\chi$PT arises due to the presence of additional momentum-dependent (derivative) interactions in $\chi$PT and the presence of electromagnetically neutral pions that interact with the charged pions via strong interactions but do not couple directly to the external magnetic field. We find that while the vortex lattice structure is hexagonal similar to vortices in GL theory, the condensation energy (relative to the normal vacuum state in a uniform, external magnetic field) is smaller (larger in magnitude) due to the presence of derivative interactions. Furthermore, we establish that neutral pions do not condense in the vortex lattice near the upper critical field.
\end{abstract}
\end{frontmatter}
\section{Introduction}
Quantum Chromodynamics (QCD) is the theory of strong interactions with a very rich phase structure that exhibits chiral symmetry breaking in the QCD vacuum and color confinement. It also has a rich structure in the presence of an external magnetic field, which is of relevance in various physical scenarios including magnetars, which can possess fields of order $10^{10}$ T~\cite{Harding:2006qn} and RHIC collisions, which involve beams of charged nuclei that produce magnetic fields of order $10^{16}$ T~\cite{Deng:2012pc}. In the presence of magnetic fields, the QCD vacuum is known to exhibit dimensional reduction~\cite{Miransky:2002rp}, whereby the pairing of quarks and antiquarks via the chiral condensate~\cite{Cohen:2007bt,Werbos:2007ym,Bali:2012zg} is enhanced (magnetic catalysis at zero temperature) and the vacuum also exhibits superconductivity through $\rho-$meson condensation~\cite{Chernodub:2012bj, Chernodub:2012bq}.

In this paper, we will study the effect of an external uniform magnetic field in the context of finite isospin three-color, two-flavor $\chi $PT. The problem (at zero external electromagnetic field) was first considered by Son and Stephanov~\cite{Son:2000by,Son:2000xc}. It was found that for chemical potentials greater than the pion mass, a condensate of pions is formed in the QCD vacuum. The $\chi$PT results are model independent and valid if the parameters involved ($x$) satisfy $x\ll\Lambda_{\rm hadr}\sim 4\pi f_{\pi}$, where $x$ can be momentum ($p$), mass ($m_{\pi}$), isospin chemical potential  ($\mu_{I}$), temperature ($T$) or an external magnetic field ($\sqrt{eH}$). Similar studies at finite isospin and finite baryon densities have been performed in QCD with different representations (of quarks) and number of colors~\cite{Kogut:2000ek} including two-color, two-flavor, which possess a rich phase structure~\cite{Splittorff:2000mm} with both pion and diquark condensation and an unusual first order transition~\cite{Adhikari:2018kzh} and QCD with arbitrary number of quarks in the adjoint representation. Furthermore, finite isospin QCD has been extended to asymptotically large isospin chemical potentials~\cite{Cohen:2015soa}, where $\chi$PT is not valid but analytical studies are still possible due to asymptotic freedom and the ability to find the BCS gap similar to large baryon chemical potentials, where QCD exhibits color-flavor-locking~\cite{Alford:1997zt,Alford:1998mk}. Most recently, finite isospin QCD has been studied using lattice methods for a wide range of chemical potentials (including intermediate ones). Additionally, there is also a model-dependent study based on the quark-meson model that incorporates quarks fluctuations (at one-loop in the on-shell renormalization scheme) with results that are largely in agreement with lattice results~\cite{Adhikari:2018cea}. 

A number of important results have been found in these most recent studies: there is a second order phase transition to a pion condensed phase for chemical potentials greater than the pion mass (in the normalization of Refs.~\cite{Son:2000by,Son:2000xc}, which we use in this paper). This transition remains largely unaffected at finite temperature up to $\sim150$ MeV. For chemical potentials larger than $\sim 1.2m_{\pi}$, the ``pion condensed" phase is sustained up to temperatures of $\sim 160$ MeV up to a chemical potential of $\sim 2m_{\pi}$ (which is the maximum chemical potential probed by lattice QCD). There is also a chiral (deconfinement) crossover transition at temperatures of $\sim160$ MeV at zero isospin, which decreases steadily with increasing chemical potentials and meets the pion phase transition line at $\mu_{I}\sim m_{\pi}$ and $T\sim151$ MeV. The quark-meson model-based results of Ref.~\cite{Adhikari:2018cea} are largely in agreement with the lattice-results except for the chiral (deconfinement) transition, which occurs at higher temperatures in the model, which may be explained by the presence of only two lightest quark flavors unlike the lattice study, which also includes the strange quark.

While lattice studies of finite isospin QCD are possible due to the absence of the fermion sign problem (unlike at finite baryon chemical potentials), the sign problem reemerges when both isospin and magnetic fields are present simultaneously due to flavor symmetry breaking in the presence of an external electromagnetic field~\cite{Endrodi:2014lja}, which leads to the loss of $\gamma_{5}$ hermicity unless the charges of the up and the down quarks are equal and opposite. As such, $\chi$PT serves as a useful effective theory in the study of the finite isospin phase diagram in external magnetic fields. Previously, in Ref.~\cite{Adhikari:2015wva} it was suggested that pions behave like type-II superconductors and the relevant single vortex structures and the critical magnetic field, $H_{c1}$, where the transition from superconducting pions to a single pionic vortex, were found. We generalize and improve upon the results by considering possible vortex lattice solutions and the corresponding condensation energy (density), relative to the normal vacuum in the presence of a uniform, external magnetic field. In this study, as in Ref.~\cite{Adhikari:2015wva}, we ignore the electromagnetic interactions between the condensed pions similar to the study of (saturated) nuclear matter (for instance, see Ref.~\cite{Adhikari:2013dfa}). The assumption is justified within the regime of validity of $\chi$PT (see Ref.~\cite{Adhikari:2015wva} for a detailed discussion) since the interactions between pions are dominated by strong interaction effects.

The paper is organized as follows: we begin in section~\ref{lagrangiansuperfluidity} with the $\chi$PT Lagrangian at finite isospin and electromagnetic fields and briefly review the resulting phase structure at zero external field. In section~\ref{type}, we summarize the argument for type-II superconductivity in finite isospin $\chi$PT with a uniform, external magnetic field that was first present in Ref.~\cite{Adhikari:2015wva} and follow that in section~\ref{magneticvortexlattice} with the computation of the magnetic vortex lattice solutions and the corresponding condensation energy (relative to the normal vacuum in an external magnetic field). In subsection~\ref{neutralpioncondensation}, we argue that neutral pions do not condense within the magnetic vortex lattice near the upper critical field and end with concluding remarks in section~\ref{conclusion}.

\section{Lagrangian and Superfluidity}
\label{lagrangiansuperfluidity}
\noindent
We begin with the Lagrangian for chiral perturbation theory at $\mathcal{O}(p^{2})$ in the presence of an isospin chemical potential and an external electromagnetic field:
\begin{equation}
\begin{split}
\mathcal{L}&=-\frac{1}{4}F_{\mu\nu}F^{\mu\nu}+\frac{f_\pi^2}{4}{\rm Tr}\left [ D_{\mu} \Sigma (D^{\mu} \Sigma)^{\dagger}\right ]\\
&+\frac{f_\pi^2 m_\pi^2}{4} {\rm Tr}\left [ \Sigma+\Sigma^{\dagger}-2\mathbb{1}\right ]\ ,
\end{split}
\end{equation}
where $\Sigma$ is an $SU(2)$ matrix, $m_{\pi}$ is the pion mass, $f_{\pi}$ is the pion decay constant, $F_{\mu\nu}$ is the electromagnetic tensor and $D_{\mu}\Sigma$ is a covariant derivative. They are defined as:
\begin{equation}
F_{\mu\nu}\equiv\partial_{\mu}A_{\nu}-\partial_{\nu}A_{\mu}\ ,
\end{equation}
\begin{equation}
\begin{split}
D_{\mu} \Sigma&\equiv\partial_{\mu}\Sigma-i [v_{\mu},\Sigma ]\\
(D_{\mu} \Sigma)^{\dagger}&=\partial_{\mu}\Sigma^{\dagger}-i [v_{\mu},\Sigma^{\dagger} ]\ ,
\end{split}
\end{equation}
\begin{equation}
v_{\mu}=\delta_{\mu 0}\mu_{I}I-eA_{\mu}Q\ ,
\end{equation}
where $A_{\mu}$ is the electromagnetic gauge field, $I$ the isospin matrix and $Q$ the charge matrix. These matrices are defined as 
\begin{equation}
\begin{split}
I&={\rm diag}\left(\frac{1}{2},-\frac{1}{2}\right)=\frac{1}{2}\tau_{3}\ ,\\
Q&={\rm diag}\left (\frac{2}{3},-\frac{1}{3}\right )=\frac{1}{6}\mathbb{1}+\frac{1}{2}\tau_{3}\ ,
\end{split}
\end{equation}
where $\mathbb{1}$ is the $2\times 2$ identity matrix and $\tau_{3}$ the third Pauli matrix. Finally, note we have added an arbitrary constant to the Lagrangian such that the normal vacuum of QCD has zero energy.
In the presence of an isospin chemical potential (and no electromagnetic fields), the system exhibits superfluidity for large enough chemical potentials~\cite{Son:2000by,Son:2000xc}. This can be easily seen using the parametrization
\begin{equation}
\Sigma=\cos\rho+i\hat{\phi}_{i}\tau_{i}\sin\rho\ ,
\end{equation}
with the condition $\hat{\phi}_{i}\hat{\phi}_{i}=1$ ensuring that the unitary condition, $\Sigma^{\dagger}\Sigma=\mathbb{1}$, is satisfied. Here $\tau_{i}$, where $i=1,2,3$, represent the Pauli matrices. The resulting static potential is
\begin{equation}
V=-\frac{f_{\pi}^{2}m_{\pi}^{2}}{2}\left [2(\cos\rho-1)+\sin^{2}\rho\left \{\frac{\mu_{I}^{2}}{m_{\pi}^{2}}(\hat{\phi}_{1}^{2}+\hat{\phi}_{2}^{2}) \} \right \} \right ]\ ,
\end{equation}
From the potential, it is straightforward to note that $\langle\hat{\phi}_{1}^{2}+\hat{\phi}_{2}^{2}\rangle=1$, i.e. $\langle\hat{\phi}_{3} \rangle=0$ (neutral pions do not condense). Furthermore, the ground state of the system is the normal vacuum, i.e. $\langle \rho \rangle=0$ if $\frac{\mu_{I}^{2}}{m_{\pi}^{2}}\le1$ and a system of charged pions, i.e. $\langle \rho \rangle=\arccos\left (\frac{m_{\pi}^{2}}{\mu_{I}^{2}}\right )$ if $\frac{\mu_{I}^{2}}{m_{\pi}^{2}}\ge1$.
\section{Type-II Superconductivity in $\chi$PT }
\label{type}
A charged superfluid in the presence of an external magnetic (which we will assume to be uniform and pointing in the positive z-direction as is standard) is expected to behave like a superconductor. If magnetic vortices (with quantized flux) are supported then the system is a type-II superconductor, otherwise type-I. A method for establishing the type of superconductivity is through the comparison of the first critical field $H_{c}$ (from a spatially uniform superconducting phase to the normal phase) with $H_{c1}$ (from the spatially, uniform superconducting phase to a single vortex). $H_{c}$ and $H_{c1}$ can be found by comparing the Gibbs free energy (densities) $\mathcal{G}$ of the relevant phases, defined as
\begin{equation}
\mathcal{G}=\mathcal{H}-\vec{M}\cdot \vec{H}\ ,
\end{equation}
where $\mathcal{H}$ is the Hamiltonian (density) assuming time-independence is, 
\begin{equation}
\begin{split}
\mathcal{H}&=\frac{1}{4}F_{ij}F^{ij}+\frac{f_{\pi}^{2}}{4}\Tr\left [D_{i}\Sigma (D_{i}\Sigma)^{\dagger}\right ]\\
&-\frac{f_{\pi}^{2}m_{\pi}^{2}}{4}\Tr\left [\Sigma+\Sigma^{\dagger}-2\mathbb{1} \right ]\ ,
\end{split}
\end{equation}
\begin{figure}
\label{singlevortex}
  \includegraphics[width=0.45\textwidth]{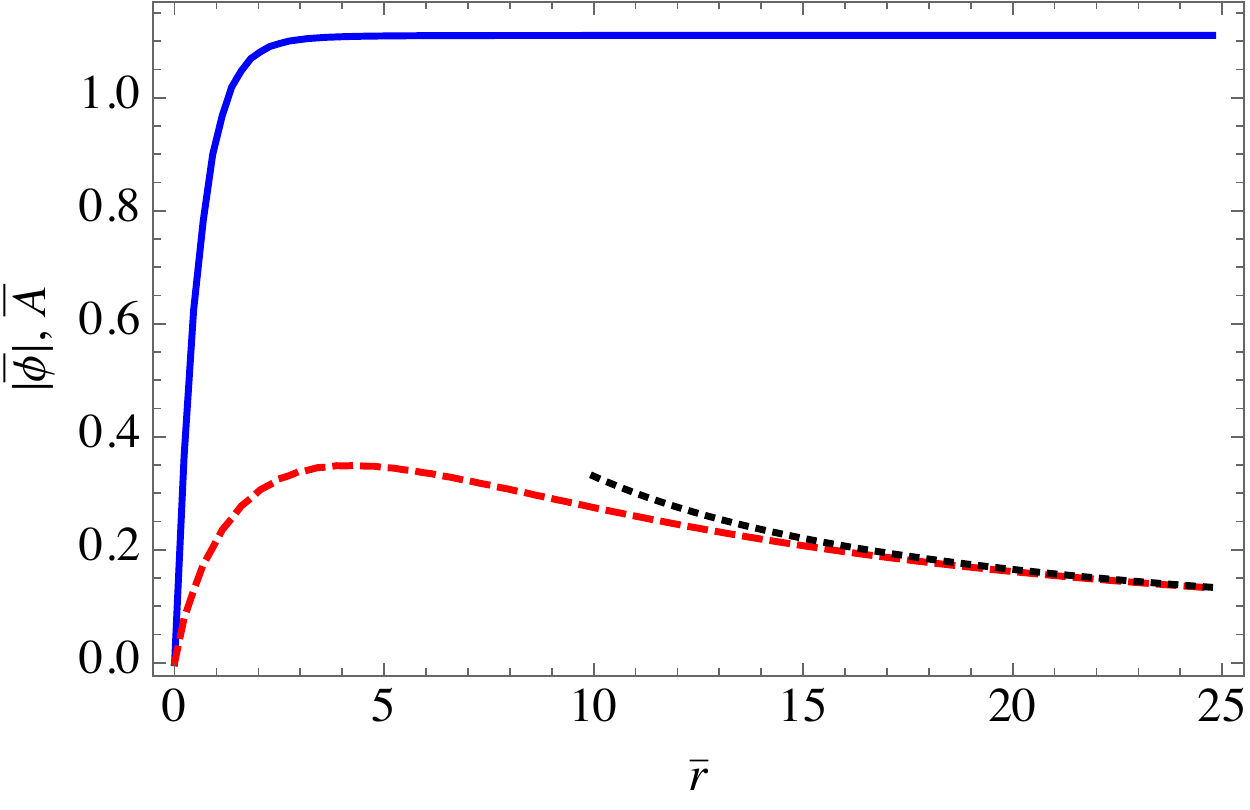}
\caption{The plot shows a single vortex at $\mu_{I}=1.5m_{\pi}$. The solid (blue) curve is the pion, the dashed (red) curve is the gauge field and the dotted (black) curve shows $\frac{1}{e\bar{r}}$, $e=\sqrt{\frac{4\pi}{137}}$. The scale is set by the pion decay constant $f_{\pi}$ with the definitions $\bar{A}=A/f_{\pi}$, $\bar{\phi}=\phi/f_{\pi}$ and $\bar{r}=f_{\pi}r$.}
\end{figure}
$\vec{M}\equiv\vec{B}-\vec{H}$ is the magnetization (density), with $\vec{B}=\vec{\nabla}\times \vec{A}$. In the uniform superconducting phase $\vec{B}=0$ and in the vortex phase, the magnetic flux $\Phi_{n}=2\pi n/e$, where $n$ is a non-zero integer. This results in the following critical magnetic field~\cite{Adhikari:2015wva} assuming type-I superconductivity
\begin{equation}
H_{c}=f_{\pi}m_{\pi}\left (\frac{\mu_{I}}{m_{\pi}}-\frac{m_{\pi}}{\mu_{I}}\right )\ ,
\end{equation}
and the following critical magnetic field assuming type-II superconductivity (from the homogeneous superconducting phase to the single vortex phase)
\begin{equation}
H_{c1}=\frac{\textrm{string tension}}{\textrm{flux}}\ ,
\end{equation}
where the string tension is the condensation energy (not free energy) per unit length (relative to the superconducting phase) of a single vortex with flux, $\Phi_{1}=2\pi/e$. If $H_{c1}<H_{c}$, then the system behaves as a type-II superconductor. The condition implies that the surface (free) energy~\cite{tinkham2004introduction} associated with  magnetic vortices is negative and thus its condensation is favored over the transition to the normal vacuum. 

\begin{figure}
\includegraphics[width=0.45\textwidth]{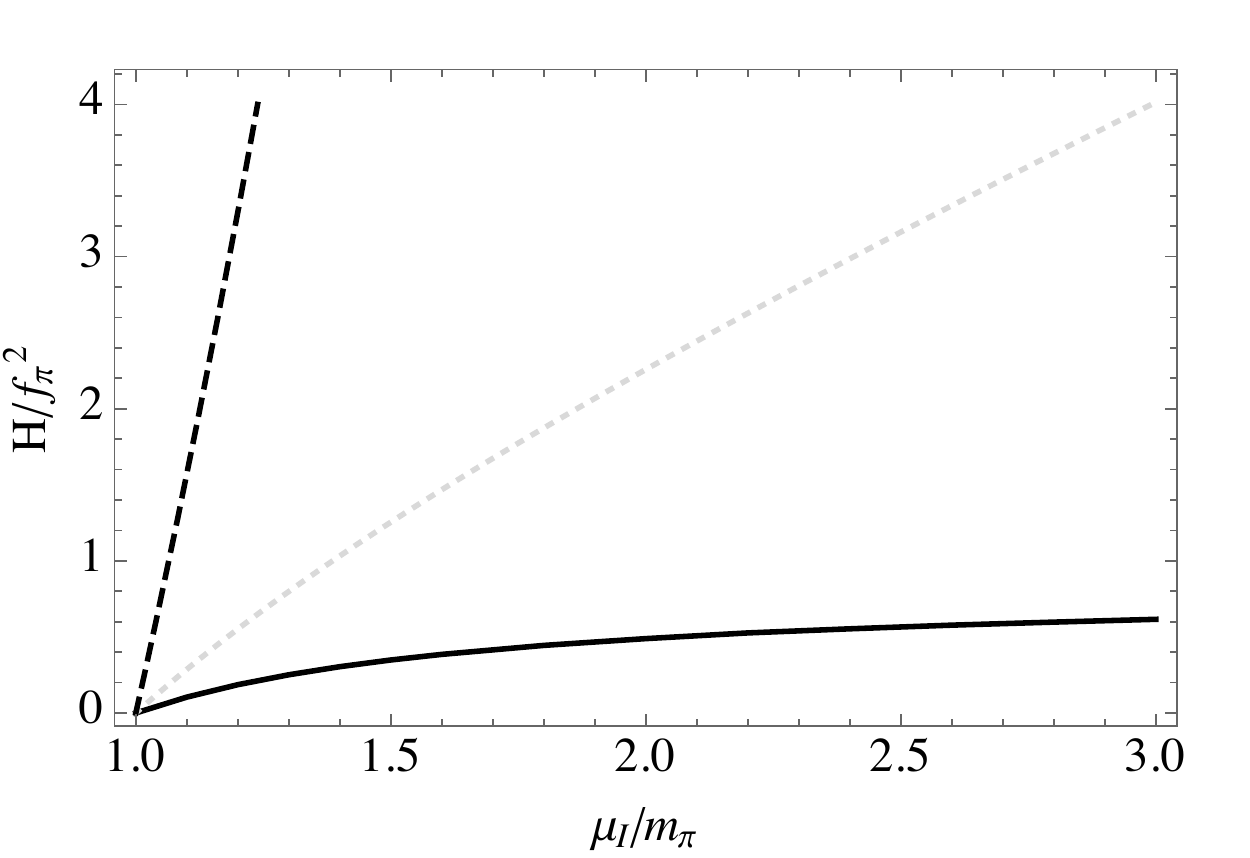}
\caption{The plot shows the phase diagram of $\chi$PT in an external magnetic field $H\equiv B_{\rm ext}$ at different chemical potentials with $\gamma\equiv\frac{\mu_{I}}{m_{\pi}}$. The solid line is a first order transition from the uniform, superconducting phase to the vortex phase and the dashed line is a second order transition from the vortex (lattice) phase to the normal vacuum. The gray line represents $H_{c}$, which is the critical field from the uniform, superconducting phase to the normal vacuum, assuming type-I superconductivity. See text for explanation.}\label{phase}
\end{figure}

In Fig.~\ref{phase}, we show the critical external field $H_{c1}$~\footnote{$H_{c1}$ is found using the discretized version of the leading order $\chi$PT Hamiltonian with discretization corrections of $\mathcal{O}(h^{2})$. The lattice step $h$ in units of $f_{\pi}$ was taken to be in the range $0.045\le h\le 0.2$ and the continuum $H_{c1}$ found by extrapolation. The box size $L$ (in units of $f_{\pi}$) was in the range $18\le L \le 30$ for $3\ge\gamma\ge 1.1$, where $\gamma\equiv\frac{\mu_{I}}{m_{\pi}}$.} from the spatially uniform, superconducting phase to a single vortex phase, which breaks translational invariance. We also show the critical external field $H_{c2}\equiv B_{c}$ from the vortex lattice phase to the normal vacuum. As the external field increases between $H_{c1}$ and $H_{c2}$ the population density of the vortices increases. Also, in Fig.~\ref{phase}, we show the critical field $H_{c}$ obtained assuming type-I superconductivity. From the diagram, we see that $H_{c}> H_{c1}$ for $\mu_{I}>m_{\pi}$ suggesting that pions in $\chi PT$ behave as type-II superconductors as pointed out in Ref.~\cite{Adhikari:2015wva}.
\section{Magnetic Vortex Lattice}
\label{magneticvortexlattice}
While the lower critical field, $H_{c1}$, was found numerically, the upper critical field $H_{c2}$ can be found using the dispersion relation of a charged pion (with charge $\pm e$), which becomes tachyonic for external magnetic fields ($H\equiv B_{\rm ext}$) larger than
\begin{equation}
\label{Bc}
H_{c2}\equiv B_{c}=\frac{\mu_{I}^{2}-m_{\pi}^{2}}{e}\ ,
\end{equation}
with $e=\sqrt{\frac{4\pi}{137}}$.
The primary objective of this section is to find the condensation energy associated with vortex lattices and argue that the vortices arrange themselves in a hexagonal lattice similar to vortex lattices in the Abelian Higgs Model.
We define the condensation energy (of the vortex lattice) as
\begin{equation}
\label{condensationenergy}
\mathcal{E}\equiv\langle\mathcal{H} \rangle-\frac{1}{2}B_{\rm ext}^{2}\ ,
\end{equation}
where $\mathcal{H}$ is the Hamiltonian density, i.e. free energy density of the vortex lattice, while $\frac{1}{2}B_{\rm ext}^{2}$ is the free energy density of the normal vacuum in a uniform external magnetic field, $B_{\rm ext}$. The vortex lattice condenses if $\mathcal{E}<0$. $\langle\mathcal{O} \rangle$ is the expectation value of the operator, $\mathcal{O}$, over the transverse plane. It is defined as
\begin{equation}
\langle\mathcal{O}\rangle=\frac{1}{A_{\perp}}\int dxdy\ \mathcal{O}\ ,
\end{equation}
where the integral is being performed over the $x-y$ (transverse) plane and $A_{\perp}$ is the area of the transverse plane.

The magnetic field $B_{\rm ext}$ is uniform and pointing in the longitudinal (i.e. positive z) direction. We will find the condensation energy in the regime 
\begin{equation}
\label{Bregime}
\frac{B_{c}-B_{\rm ext}}{B_{c}}\ll 1\ ,
\end{equation}
using Abrikosov's method (of successive approximations).
Using the non-linear parametrization for $\Sigma$,
\begin{equation}
\begin{split}
\Sigma&=\exp\left (i\frac{\phi}{f_{\pi}} \right )\ ,
\end{split}
\end{equation}
where 
\begin{equation}
\begin{split}
\phi=\phi_{i}\tau_{i}&=
\begin{pmatrix} 
\phi_{3}& \phi_{1}-i\phi_{2}\\
\phi_{1}+i\phi_{2}& \phi^{3}
\end{pmatrix}\\
&\equiv
\begin{pmatrix} 
\pi_{0}& \sqrt{2}\pi_{+}\\
\sqrt{2}\pi_{-} & \pi_{0}
\end{pmatrix}\ ,
\end{split}
\end{equation}
we can expand the $\mathcal{O}(p^{4})$ Lagrangian near the critical field $B_{c}$ assuming $|\pi_{+}|\gtrsim 0$ in the regime defined in Eq.(~\ref{Bregime})
\begin{equation}
\mathcal{L}\equiv\mathcal{L}_{4\pi}+\delta \mathcal{L}\ ,
\end{equation}
where $\mathcal{L}_{4\pi}$ are contributions to the $\chi$PT Lagrangian with four or fewer pion fields (that contribute at $\mathcal{O}(p^{2})$) and $\delta\mathcal{L}$ contains all other terms. (Single Vortices in theories with derivatives interactions were first studied in Ref.~\cite{Babichev:2007tn}
$\mathcal{L}_{4\pi}$ is
\begin{equation}
\begin{split}
\mathcal{L}_{4\pi}&=-\frac{1}{4}F_{\mu\nu}F^{\mu\nu}+i\mu_{I}(\pi_{-}\partial_{0}\pi_{+}-\pi_{+}\partial_{0}\pi_{-}  )\\
&-i\frac{2\mu_{I}}{3f_{\pi}^{2}}\pi_{-}\pi_{+}\left (\pi_{-}\partial_{0}\pi_{+}-\pi_{+}\partial_{0}\pi_{-} \right )\\
&+D_{\mu}\pi_{+}D^{\mu\dagger}\pi_{-}+(\mu_{I}^{2}-m_{\pi}^{2})\pi_{-}\pi_{+}\\
&-\frac{4\mu_{I}^{2}-m_{\pi}^{2}}{6f_{\pi}^{2}}\pi_{+}^{2}\pi_{-}^{2}-\frac{1}{3f_{\pi}^{2}}\pi_{-}\pi_{+}D_{\mu}\pi_{+}D^{\mu\dagger}\pi_{-}\\
&+\frac{1}{6f_{\pi}^{2}}\left [\pi_{+}^{2}D_{\mu}^{\dagger}\pi_{-}D^{\mu\dagger}\pi_{-}+\pi_{-}^{2}D_{\mu}\pi_{+}D^{\mu}\pi_{+} \right ]\\
&+\frac{1}{2}\partial_{\mu}\pi_{0}\partial^{\mu}\pi_{0}-\frac{1}{2}m_{\pi}^{2}\pi_{0}^{2}\\
&+\frac{m_{\pi}^{2}}{24f_{\pi}^{2}}\pi_{0}^{4}+\frac{m_{\pi}^{2}}{2f_{\pi}^{2}}\pi_{0}^{2}\pi_{-}\pi_{+}\ ,
\end{split}
\end{equation}
where 
\begin{equation}
\begin{split}
D_{\mu}\pi_{+}&=(\partial_{\mu}+ieA_{\mu})\pi_{+}\\
D^{\dagger}_{\mu}\pi_{-}&=(\partial_{\mu}-ieA_{\mu})\pi_{-}\ .
\end{split}
\end{equation}
We proceed by assuming that $\pi_{0}=0$. However, this assumption needs to be justified within $\chi$PT since neutral particles are known to condense in vortex phases~\cite{Adhikari:2017oxb}. We will discuss the issue of neutral pion condensation further in section~\ref{neutralpioncondensation}. The equation of motion of $\mathcal{L}_{4\pi}$ for $\pi_{+}$ ignoring $\pi_{0}$ is
\begin{equation}
\begin{split}
&D_{\mu}D^{\mu}\pi_{+}-(\mu_{I}^{2}-m_{\pi}^{2})\pi_{+}+\frac{4\mu_{I}^{2}-m_{\pi}^{2}}{3f_{\pi}^{2}}(\pi_{-}\pi_{+})\pi_{+}\\
&-\frac{1}{3f_{\pi}^{2}}D_{\mu}\left [(D^{\mu}\pi_{+})\pi_{-}\pi_{+} \right ]+\frac{1}{3f_{\pi}^{2}}D_{\mu}\left[(D^{\mu\dagger}\pi_{-})\pi_{+}\pi_{+}\right]\\
&+\frac{1}{3f_{\pi}^{2}}(D_{\mu}^{\dagger}\pi_{-})(D^{\mu}\pi_{+})\pi_{+}-\frac{1}{3f_{\pi}^{2}}(D_{\mu}\pi_{+})(D^{\mu}\pi_{+})\pi_{-}=0\ .
\end{split}
\end{equation}
and that of the electromagnetic field is
\begin{equation}
\partial_{\mu}F^{\mu\nu}=-j^{\nu}\ ,
\end{equation}
where the electromagnetic current is
\begin{equation}
\begin{split}
j^{\nu}&=-ie\left [\pi_{-}D^{\mu}\pi_{+}-(D^{\mu\dagger}\pi_{-})\pi_{+} \right ]\left[1-\frac{2}{3f_{\pi}^{2}}\pi_{-}\pi_{+}\right ]\ .
\end{split}
\end{equation}
The equation of motion for the electromagnetic fields is unmodified by the presence of neutral pions.

Assuming time-independence and the absence of electric fields, i.e. $A_{0}=0$~\cite{Adhikari:2015wva}, we can use the equations of motion for $\pi_{+}$ to rewrite the Hamiltonian containing up to four fields.
\begin{equation}
\label{H4}
\begin{split}
\mathcal{H}_{4\pi}&=\frac{1}{2}B^{2}+\frac{f_{\pi}^{2}}{4}D_{i}\pi_{-}D_{i}\pi_{+}-\frac{4\mu_{I}^{2}-m_{\pi}^{2}}{6f_{\pi}^{2}}(\pi_{-}\pi_{+})^{2}\\
&+\frac{1}{3f_{\pi}^{2}}(D_{i}^{\dagger}\pi_{-})(D_{i}\pi_{+})\pi_{-}\pi_{+}\\\
&-\frac{1}{6f_{\pi}^{2}}\left [(D_{i}^{\dagger}\pi_{-})^{2}\pi_{+}^{2}+(D_{i}\pi_{+})^{2}\pi_{-}^{2} \right ]+\cdots\ ,\\ 
\end{split}
\end{equation}	
where we have ignored terms with more than four scalar fields in the $\mathcal{O}(p^{2})$ Lagrangian. Note that in the absence of derivative interactions in $\mathcal{H}_{4\pi}$ above reduces to the standard Abelian Higgs Model result~\cite{Chernodub:2011gs}. 

We use the following notation in what follows:
\begin{equation}
\begin{split}
z&=x+iy,\ \overline{z}=x-iy\\
\partial&=\frac{1}{2}(\partial_{x}-i\partial_{y}),\ \overline{\partial}=\frac{1}{2}(\partial_{x}+i\partial_{y})\ .
\end{split}
\end{equation}

We proceed by using the method of successive approximations to find the vortex lattice solution and the condensation (free) energy relative to the normal vacuum (which has a free energy of $\frac{1}{2}B_{\rm ext}^{2}$). The method is a perturbative expansion with the following power counting scheme
\begin{equation}
\begin{split}
B&=B_{0}+\epsilon^{2} \delta B\\
\pi_{+}&=\epsilon\tilde{\pi}_{+}+\epsilon^{3}\delta\pi_{+}
\end{split}
\end{equation}
At leading order, i.e. $\mathcal{O}(\epsilon)$, within the method of successive approximations, $B_{0}=B_{c}$ (see Eq.~(\ref{Bc})), and $\tilde{\pi}_{+}$ satisfies the following linearized equation of motion 
 
\begin{equation}
\label{eompi}
\left (2\overline{\partial}+\frac{eB_{0}}{2}z \right )\tilde{\pi}_{+}=0\ ,
\end{equation}
the solutions of which are the Landau level wave functions, which in the symmetric gauge assume the following form~\cite{Chernodub:2011gs,Abrikosov:1956sx}.
\begin{equation}
\label{piplus}
\begin{split}
\tilde{\pi}_{+}&=\sum_{n=-\infty}^{\infty}C_{n}\phi_{n}(\nu,z,\bar{z})\\
\phi_{n}(\nu,z,\bar{z})&=e^{-\pi\nu^{2}n^{2}-\frac{\pi}{2L_{B_{0}}}\left (|z|^{2}+z^{2} \right)+\frac{2\pi}{L_{B_{0}}}\nu n z}\ ,
\end{split}
\end{equation}
where $L_{B_{0}}=\sqrt{\frac{2\pi}{B_{0}}}$. It is standard to consider the most symmetric solutions, i.e. periodic lattice solutions, which requires choosing $C_{n}=C_{n+N}$. (For $N=1$, the solution is a square lattice, for $N=2$ it is a hexagonal lattice and $N=3$ it is a paralleogrammic lattice. For further possibilities and details, see Ref.~\cite{Chernodub:2011gs}) We note that for $B=B_{c}$, $|\pi_{+}|=0$. For fields $B_{\rm ext}\lesssim B_{c}$, we expect $|\pi_{+}|\gtrsim 0$. In order to find the condensation energy and the charged pion field $\pi_{+}$ below the critical field, we need to work at next to leading order, i.e. $\mathcal{O}(\epsilon^{3})$. The equation of motion for $B$ at $\mathcal{O}(\epsilon^{3})$ can be written in terms of $F^{12}\equiv -B$ (the other components of $F_{ij}$ are zero), in the following form 
\begin{equation}
\label{Bequation}
\overline{\partial} B=\frac{j^{2}-i j^{1}}{2}=-e\overline{\partial}(\tilde{\pi}_{-}\tilde{\pi}_{+})\left [1-\frac{2}{3f_{\pi}^{2}}\tilde{\pi}_{-}\tilde{\pi}_{+} \right ]\ ,
\end{equation}
Solving Eq.~(\ref{Bequation}), we find that $B$ is given by
\begin{equation}
\label{BB}
\begin{split}
B&=B_{\rm ext}-e\left ( \tilde{\pi}_{-}\tilde{\pi}_{+}-\langle\tilde{\pi}_{-}\tilde{\pi}_{+} \rangle\right )\\
&+\frac{e}{3f_{\pi}^{2}}\left ( (\tilde{\pi}_{-}\tilde{\pi}_{+})^{2}-\langle(\tilde{\pi}_{-}\tilde{\pi}_{+})^{2}\rangle\right )\\
\delta B&=B-B_{c}\ ,
\end{split}
\end{equation}
where an arbitrary constant that arises in solving the first order Eq.~(\ref{Bequation}) is fixed by the conservation of magnetic flux, i.e. $\langle B \rangle=B_{\rm ext}$~\cite{Chernodub:2011gs}. Similarly, the next to leading order, i.e. $\mathcal{O}(\epsilon^{3})$, equation of motion (For details of the calculation, see Ref.~\cite{Adhikari:2018tvz}) for the charged pion field, $\pi_{+}$, gives rise to following condition
\begin{equation}
\begin{split}
\langle(\tilde{\pi}_{-}\tilde{\pi}_{+})^{2}\rangle&=\frac{e(B_{c}-B_{\rm ext})\langle\tilde{\pi}_{-}\tilde{\pi}_{+} \rangle-e\langle\tilde{\pi}_{-}\tilde{\pi}_{+}\rangle^{2}}{\frac{2\mu_{I}^{2}+m_{\pi}^{2}}{3f_{\pi}^{2}}-e^{2}}\\
&+\mathcal{O}((\tilde{\pi}_{-}\tilde{\pi}_{+})^3)\ .
\end{split}
\end{equation}
Since our goal is to find the condensation energy at $\mathcal{O}(\epsilon^{4})$, we need to evaluate the expectation value of the Hamiltonian density in Eq.~(\ref{H4}). The difference from the Abelian Higgs model is the presence of derivative interactions. The expectation value of the first derivative interaction term can be simplified using the equations of motion at $\mathcal{O}(\epsilon^{4})$:
\begin{equation}
\begin{split}
\langle (D_{i}^{\dagger}\tilde{\pi}_{-})(D_{i}\tilde{\pi}_{+})\tilde{\pi}_{-}\tilde{\pi}_{+}\rangle&=(\mu_{I}^{2}-m_{\pi}^{2})\langle (\tilde{\pi}_{-}\tilde{\pi}_{+})^{2}\rangle\\
&+\mathcal{O}((\tilde{\pi}_{-}\tilde{\pi}_{+})^3)\\
\end{split}
\end{equation}
and the expectation value of the second derivative interaction term vanishes at $\mathcal{O}(\epsilon^{4})$ in a uniform, external magnetic field due to Eq.~(\ref{eompi}), i.e.
\begin{equation}
\begin{split}
&\langle(D_{i}^{\dagger}\tilde{\pi}_{-})^{2}\tilde{\pi}_{+}^{2}+(D_{i}\tilde{\pi}_{+})^{2}\tilde{\pi}_{-}^{2} \rangle=\mathcal{O}((\tilde{\pi}_{-}\tilde{\pi}_{+})^3)\ .
\end{split}
\end{equation}
Using the above expectation values and the result of Eq.~(\ref{BB}) in Eq.~(\ref{H4}), the expectation value of the Hamiltonian density over the transverse plane valid at $\mathcal{O}(\epsilon^{4})$ near the upper critical field ($B_{c}$) is
\begin{equation}
\label{HH}
\begin{split}
\langle\mathcal{H}\rangle&=\frac{1}{2}B_{\rm ext}^{2}-e(B_{c}-B_{\rm ext})\langle\tilde{\pi}_{-}\tilde{\pi}_{+}\rangle+\frac{e^{2}}{2} \langle \tilde{\pi}_{-}\tilde{\pi}_{+}\rangle^{2}\\
&+\frac{1}{2}\left (\frac{2\mu_{I}^{2}+m_{\pi}^{2}}{3f_{\pi}^{2}}-e^{2}\right )\langle(\tilde{\pi}_{-}\tilde{\pi}_{+})^{2}\rangle+\mathcal{O}((\tilde{\pi}_{-}\tilde{\pi}_{+})^3)\ .
\end{split}
\end{equation}
This expectation value reduces to that of the Abelian Higgs model of Ref.~\cite{Chernodub:2011gs} in the absence of derivative interactions, which is easily checked by removing the derivative interactions in Eq.~(\ref{H4}). 

By using the definition of the Abrikosov ratio~\cite{Chernodub:2011gs, Abrikosov:1956sx,abrikosov1988fundamentals},
\begin{equation}
\beta_{A}\equiv\frac{\langle (\tilde{\pi}_{-}\tilde{\pi}_{+})^{2}\rangle}{\langle\tilde{\pi}_{-}\tilde{\pi}_{+} \rangle^{2}}\ ,
\end{equation}
we can write $\langle\mathcal{H}\rangle$ in terms $\langle\pi_{+}\pi_{-} \rangle$ and $\beta_{A}$. Then,
minimizing $\langle\mathcal{H} \rangle$ with respect to $\langle\pi_{-}\pi_{+} \rangle$ (subsequently, we will minimize with respect to $\beta_{A}$), gives up to $\mathcal{O}(\epsilon^{4})$:
\begin{equation}
\begin{split}
\frac{\partial\langle\mathcal{H}\rangle}{\partial\langle\tilde{\pi}_{-}\tilde{\pi}_{+}\rangle}=0\implies\langle \tilde{\pi}_{-}\tilde{\pi}_{+}\rangle=\frac{e(B_{c}-B_{\rm ext})}{\beta_{A}\left(\frac{2\mu_{I}^{2}+m_{\pi}^{2}-e^{2}}{3f_{\pi}^{2}} \right)+e^{2}}\ .
\end{split}
\end{equation}
Plugging this result into Eq.~(\ref{HH}) and then into Eq.~(\ref{condensationenergy}), we get the following expression for the vortex lattice condensation energy (relative to the normal vacuum in a uniform magnetic field):
\begin{equation}
\mathcal{E}=-\frac{e^{2}(B_{c}-B_{\rm ext})^{2}}{2e^{2}+\beta_{A}\left (\frac{2\mu_{I}^{2}+m_{\pi}^{2}}{3f_{\pi}^{2}}-e^{2}\right )}+\mathcal{O}\left ((B_{c}-B_{\rm ext})^{4} \right )\ ,
\end{equation}
which is valid at or below the critical field $B_{c}$, above which $\mathcal{E}=0$. It is evident from the above expression that the condensation energy is minimized by the smallest possible $\beta_{A}$, which is assumed by a hexagonal (triangular) lattice structure with~\cite{abrikosov1988fundamentals,kleiner1964bulk}
\begin{equation}
\begin{split}
\beta_{A}&=1.159595\dots\\
\nu&=\frac{\sqrt[4]{3}}{\sqrt{2}}\ .
\end{split}
\end{equation}
Furthermore as discussed in full detail in Ref.~~\cite{Chernodub:2011gs}, in Eq.~(\ref{piplus}), we set
\begin{equation}
\label{CCC}
\begin{split}
C_{n}&=C_{n+2}\\
\textrm{with }C_{0}&\equiv C\textrm{ and }C_{1}=iC\ .
\end{split}
\end{equation}
\begin{figure}
\includegraphics[width=0.45\textwidth]{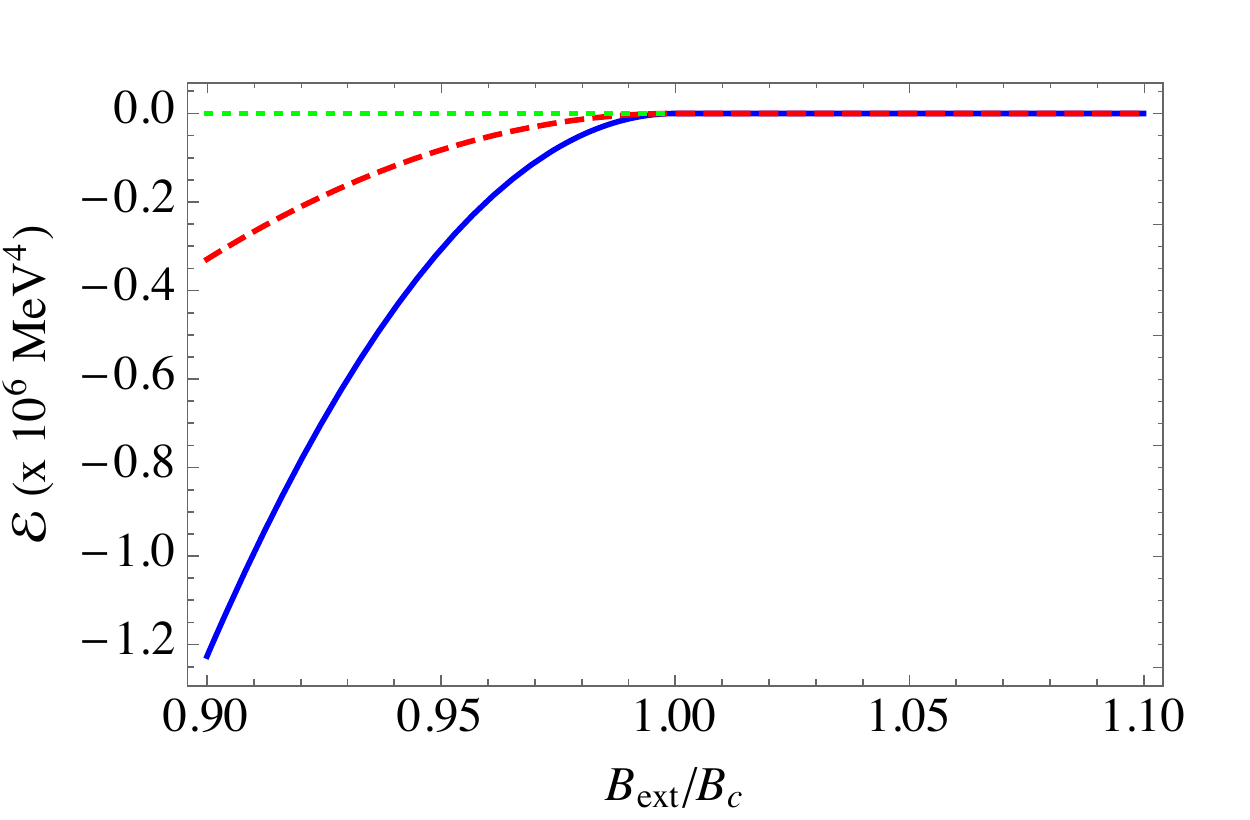}
\caption{The plot shows the condensation energy of the vortex lattice for three different values of the isospin chemical potential $\left (\gamma\equiv\frac{\mu_{I}}{m_{\pi}}\right )$: $\gamma=1$ (green, dotted), $\gamma=1.25$ (dashed) and $\gamma=1.50$ (blue, solid).}
\label{condensationenergyplot}
\end{figure}
\begin{figure*}[t!]
\centering
\includegraphics[width=0.31\textwidth]{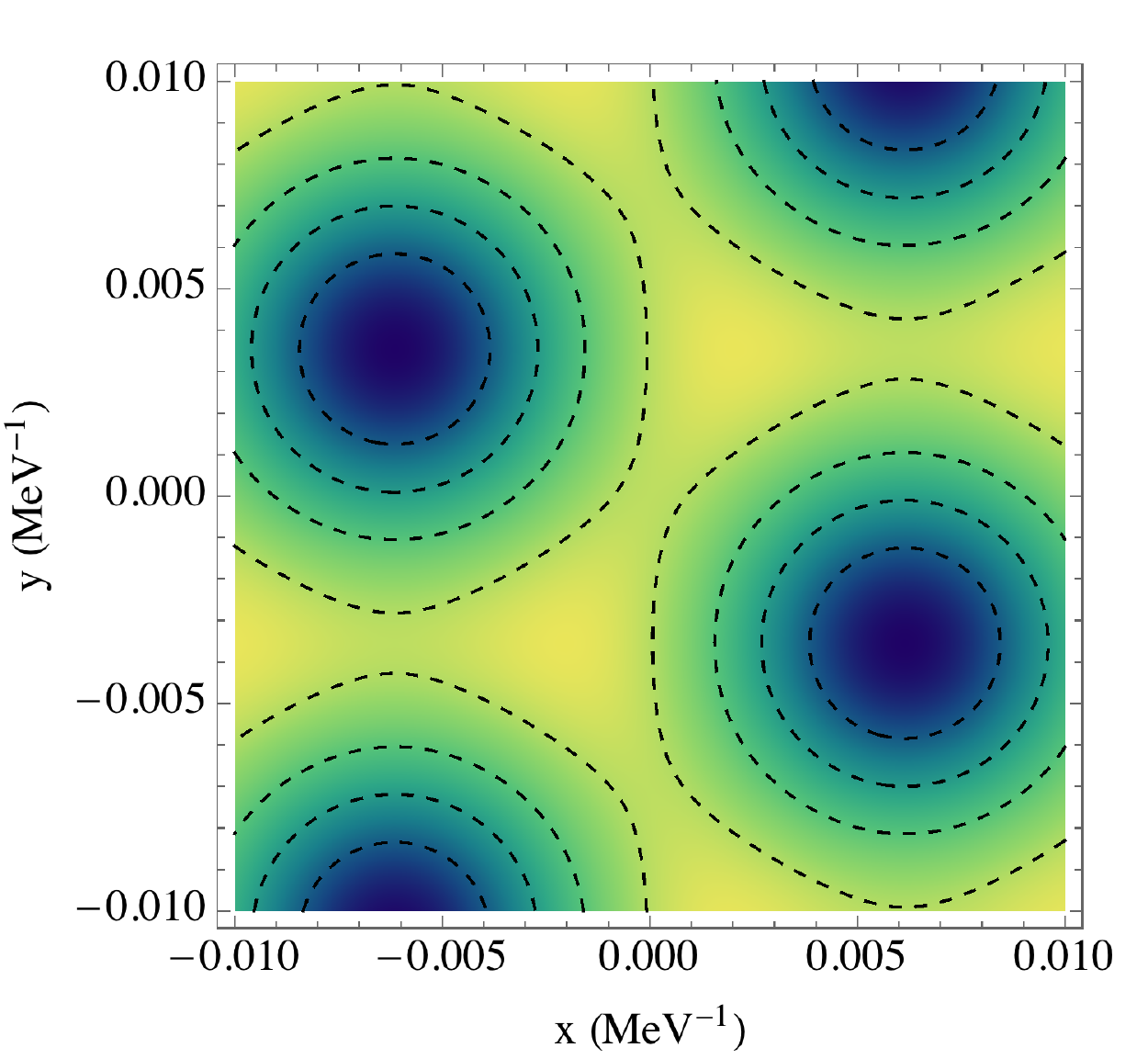}\,\,\,\,\,
\includegraphics[width=0.31\textwidth]{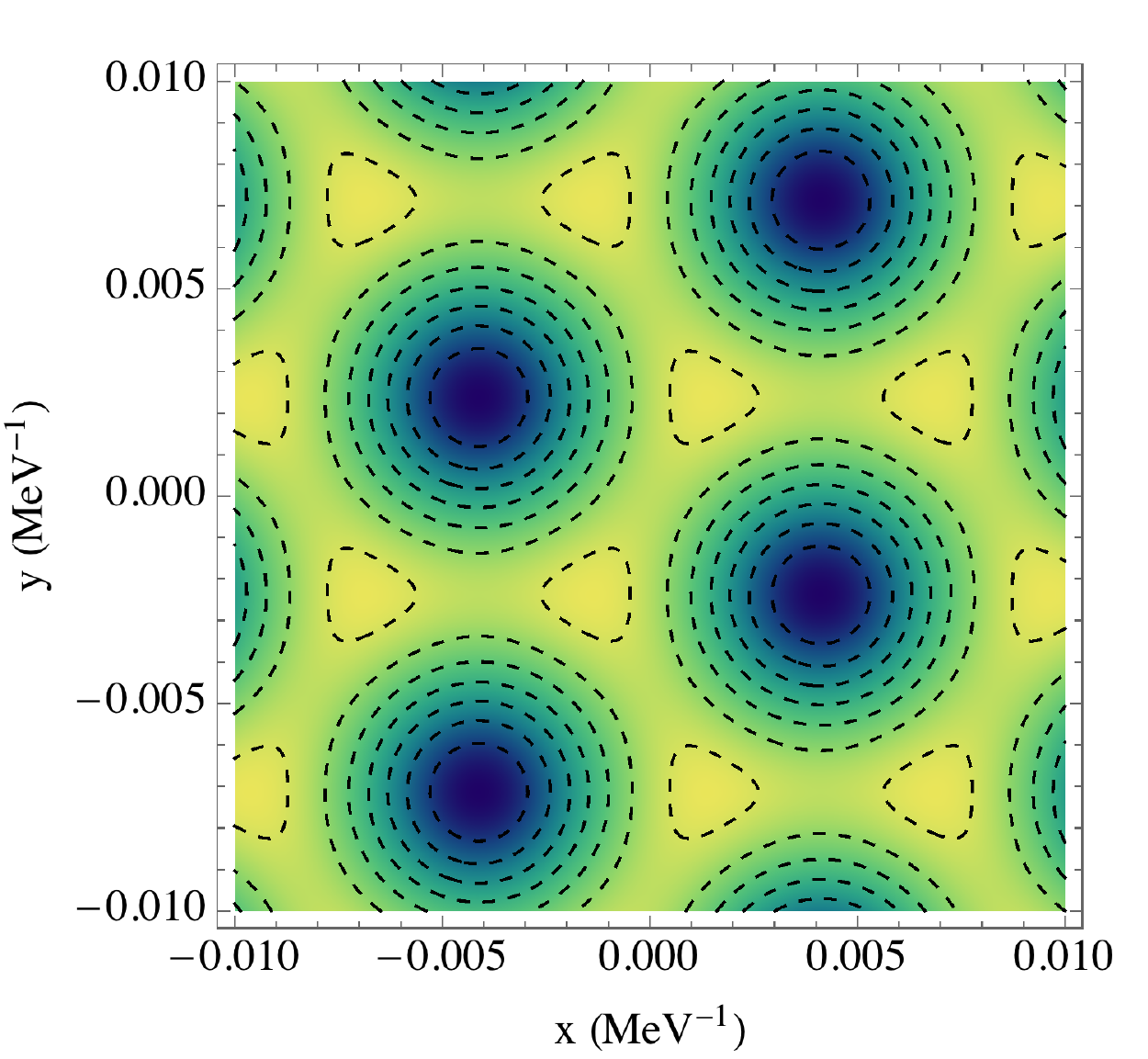}\,\,\,\,\,
\includegraphics[width=0.31\textwidth]{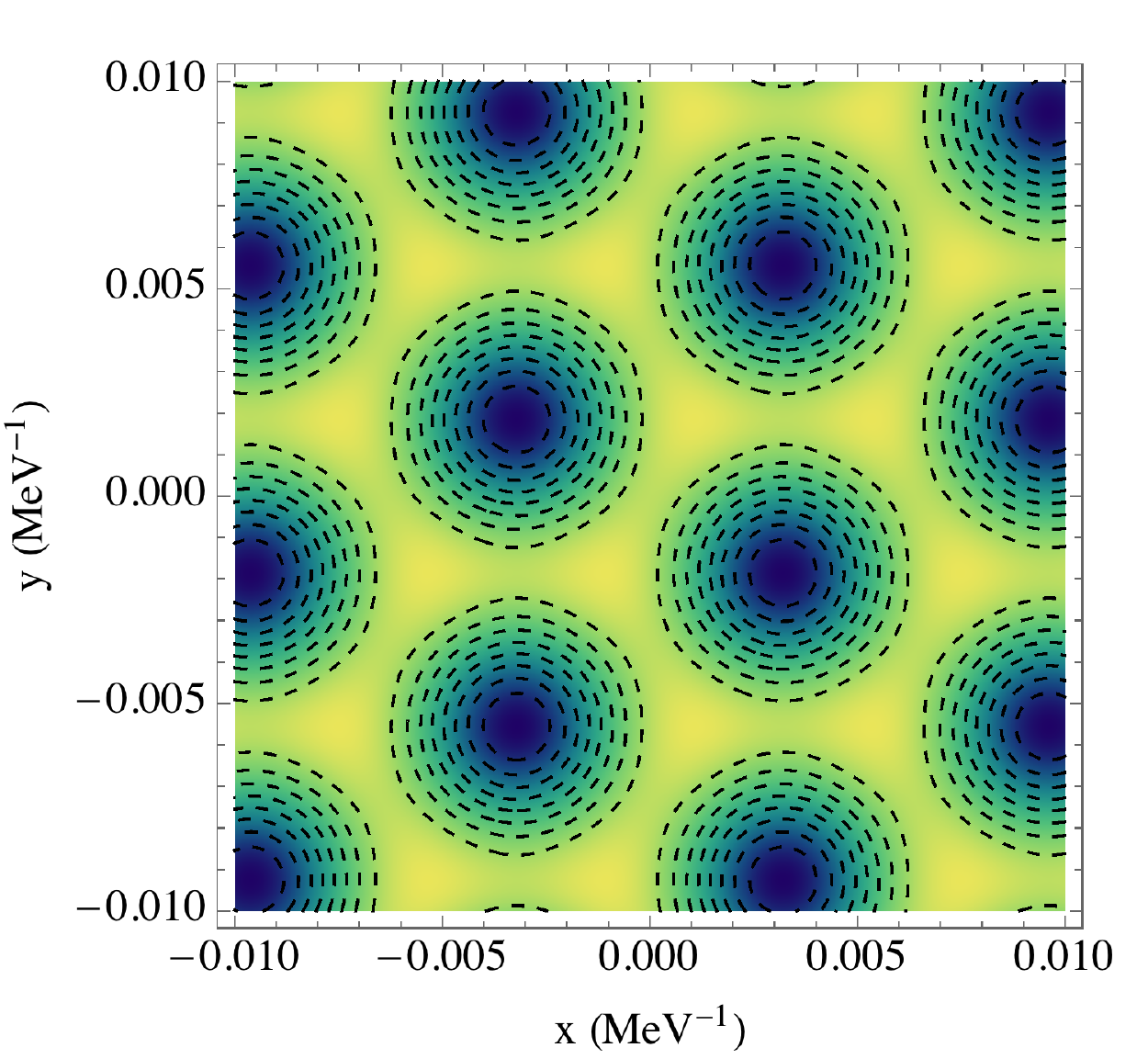}
\caption{Density plots ($|\pi_{+}|$) of vortex lattices in a square region of size $0.02\times0.02$ $\textrm{MeV}^{-2}$ on the transverse plane (x-y) for three different values of the isospin chemical potential. From left to right, $\gamma\left (\equiv\frac{\mu_{I}}{m_{\pi}}\right)=1.25,\ 1.5,\ 1.75$, for which the values of $\left | C \right |^{2}$(see Eq.~(\ref{C})) are $40.32$, $65.68$ and $84.01$ (to 4 sf) respectively. The contour lines connect regions with equal isospin density and have the same normalization across all three plots. The color shades, on the other hand, are normalized separately for each plot.} 
\label{fig:vlattice}     
\end{figure*}

In Fig~\ref{condensationenergyplot}, we show the vortex lattice condensation energy as a function of $B_{\rm ext}/B_{c}$. The vortex lattice becomes the normal vacuum when $B_{\rm ext}=B_{c}$ and the condensation energy becomes zero thereafter. As $B_{\rm ext}$ decreases below $B_{c}$, $|\pi_{+}|$ increases. Furthermore, with increasing isospin chemical potential, the superfluid pion density increases and the vortex lattices have a more negative condensation energy. The vortex lattices also have a smaller (larger in magnitude) condensation energy than they would in the absence of derivative interactions.

For an external magnetic field, $B_{\rm ext}\lesssim B_{c}$, the vortex lattice solution ~\cite{Chernodub:2011gs} is modified (with $B_{c}\rightarrow B_{\rm ext}$) such that it satisfies the flux quantization condition~\cite{Zak:1964zz,avron1978separation}
\begin{equation}
B_{\rm ext}\ \left |\vec{d}_{1}\times\vec{d}_{2}\right |=\frac{2\pi n}{e}\ ,
\end{equation}
where $n=1$ in each unit cell. In the symmetric gauge (for lattice solutions in non-symmetric gauges, see Refs.~\cite{Abrikosov:1956sx,abrikosov1988fundamentals,rosenstein2010ginzburg})
\begin{equation}
A_{x}=-\frac{B_{\rm ext}}{2}y,\ A_{y}=\frac{B_{\rm ext}}{2}x
\end{equation}
can be written using Eqs.~(\ref{piplus}) and (\ref{CCC}) in terms of the third Jacobi (elliptic) theta function $\theta_{3}$, a variational parameter~\cite{Chernodub:2011gs} which was found to be $\nu=\frac{\sqrt[4]{3}}{\sqrt{2}}$ for the hexagonal lattice and the modified magnetic length $L_{B}=\sqrt{\frac{2\pi}{eB_{\rm ext}}}$ with the following result for $\tilde{\pi}_{+}$:
\begin{equation}
\begin{split}
\tilde{\pi}_{+}&=\frac{C}{2\nu}\left [ e^{-\frac{\pi y(-i x+y)}{L_{B}^{2}}}\theta_{3}\left(\frac{-\pi(x+iy)}{2L_{B}\nu},e^{-\frac{\pi}{4\nu^{2}}} \right)\right.\\
&\left.+ie^{-\frac{\pi y(-ix+y)}{L_{B}^{2}}}\theta_{3}\left(\frac{-\pi (x+i y-L_{B}\nu)}{2L_{B}\nu},e^{-\frac{\pi}{4\nu^{2}}} \right)\right ]\ .
\end{split}
\end{equation}
We find using the above expression that~\cite{Chernodub:2011gs}
\begin{equation}
\begin{split}
\langle\tilde{\pi}_{-}\tilde{\pi}_{+} \rangle&=\frac{|C|^{2}}{\sqrt[4]{3}}\\
\langle(\tilde{\pi}_{-}\tilde{\pi}_{+})^{2} \rangle&=\frac{\beta_{A}|C|^{4}}{\sqrt{3}}
\end{split}
\end{equation}
which is independent of $B_{\rm ext}$. Plugging these into Eq.~(\ref{HH}) and minimizing with respect to $|C|$ results in the following expression for $|C|$
\begin{equation}
\begin{split}
\label{C}
&|C|=\sqrt{\frac{3^{\frac{1}{4}}e(B_{c}-B_{\rm ext})}{\beta_{A}\left (\frac{2\mu_{I}^{2}+m_{\pi}^{2}}{3f_{\pi}^{2}}
-e^{2}\right )+e^{2}}}\ ,{\rm\ for\ }B_{\rm ext}\le B_{c}\\
&|C|=0\ ,{\rm\ for\ }B_{\rm ext}\ge B_{c}\ .
\end{split}
\end{equation}
Note that, $|C|\rightarrow 0$ as $B_{\rm ext}\rightarrow B_{c}$ as expected. Also, $\tilde{\pi}_{+}$ is a periodic function with lattice vectors $\vec{d}_{1}=\frac{L_{B}}{\nu}(0,1)$ and $\vec{d}_{2}=\frac{L_{B}}{\nu}\left(\frac{\sqrt{3}}{2},\frac{1}{2}\right)$ that lie on the transverse plane.

We plot the vortex lattices, in particular ($\tilde{\pi}_{-}\tilde{\pi}_{+}$), in Fig.~\ref{fig:vlattice} for three different chemical potentials: $\gamma\equiv\frac{\mu_{I}}{m_{\pi}}=1.25,\ 1.50,\ 1.75$. As the isospin chemical potential increases, the spatial density of vortices also increases. Furthermore, the size of the pion condensate (see Eq.~(\ref{C})) increases. 
\subsection{Absence of Neutral Pion Condensation}
\label{neutralpioncondensation}
We have so far assumed that $\pi_{0}=0$, which needs to be justified~\cite{Adhikari:2017oxb}. In order to check the validity of this assumption, we consider small fluctuations of $\pi_{0}$ around its assumed value of zero~\cite{coleman1988aspects}. These fluctuations can be written in the following form:
\begin{equation}
\pi_{0}(t,\vec{x})=\sum_{n}e^{iE_{n}t}\tilde{\pi}_{0}(\vec{x})\ .
\end{equation}
Using the equation of motion for $\pi_{0}$, we find that the linearized equation (``Schr\"{o}dinger equation") for the fluctuations in $\pi_{0}$ is:
\begin{equation}
\left(\partial_{i}\partial^{i}+m_{\pi}^{2}-\frac{m_{\pi}^{2}}{f_{\pi}^{2}}\left |\tilde{\pi}_{+}\right |^{2}\right)\tilde{\pi}_{0}=E_{n}^{2}\tilde{\pi}_{0}
\end{equation}
While we are not aware of any analytical solutions to this equation, we can solve the equation using finite element methods in a unit cell (a parallelogram defined by $\vec{d}_{1}$ and $\vec{d}_{2}$) by imposing periodic boundary conditions. The resulting dispersion relation is
\begin{equation}
E_{0}^{2}=p_{i}^{2}+m_{\pi}^{2}+\delta\ ,
\end{equation}
where $\delta$ is the ground state ``energy" of the periodic potential, $-\frac{m_{\pi}^{2}}{f_{\pi}^{2}}|\tilde{\pi}_{+}|^{2}$. It depends on the external magnetic field $B_{\rm ext}$ and the isospin chemical potential $\mu_{I}$. We note that $\delta\le 0$ (see Table~\ref{delta} for some numerical values), which can be argued variationally using a uniform distribution for $\tilde{\pi}_{+}$ that trivially satisfies periodic boundary conditions, and $\delta\ge -\frac{m_{\pi}^{2}}{f_{\pi}^{2}}|C|^{2}$. Our results suggest that within the regime of validity of $\chi$PT and the method of successive approximation used in this work, the dispersion relation never becomes tachyonic: there is no neutral pion condensation in the vortex lattice near the upper critical field. Therefore, the phase diagram for type-II superconductivity of as suggested by Fig.~\ref{phase} is consistent at (or near) the two critical fields. Our calculations, however, do not preclude neutral pion condensation within magnetic vortices entirely. (see Fig.~\ref{conjecture})
\begin{center}
\begin{table}
\begin{tabular}{ |c|c|c|c|c|c|c| }
  \hline
  $B_{\rm ext}, \gamma$ & 1.1 & 1.5 & 2.0 \\ \hline
  0.99 $B_{c}$ & -31.2906  & -115.599 & -169.342 \\ \hline
  0.95 $B_{c}$ & -156.487 & -578.072 & -846.78\\ \hline
\end{tabular}
\begin{tabular}{ |c|c|c|c|c|c|c| }
  \hline
  $B_{\rm ext}, \gamma$ & 2.5 & 3.0\\ \hline
  0.99 $B_{c}$  & -197.442 & -213.695\\ \hline
  0.95 $B_{c}$ & -987.265 &-1068.51\\ \hline
\end{tabular}
\caption{The ``gap" ($\delta$) in ${\rm MeV}^{2}$  for different values of $B_{\rm ext}$ and $\gamma\equiv\frac{\mu_{I}}{m_{\pi}}$.}
\label{delta}
\end{table}
\end{center}
\section{Conclusion}
\label{conclusion}
\begin{figure}
\includegraphics[width=0.45\textwidth]{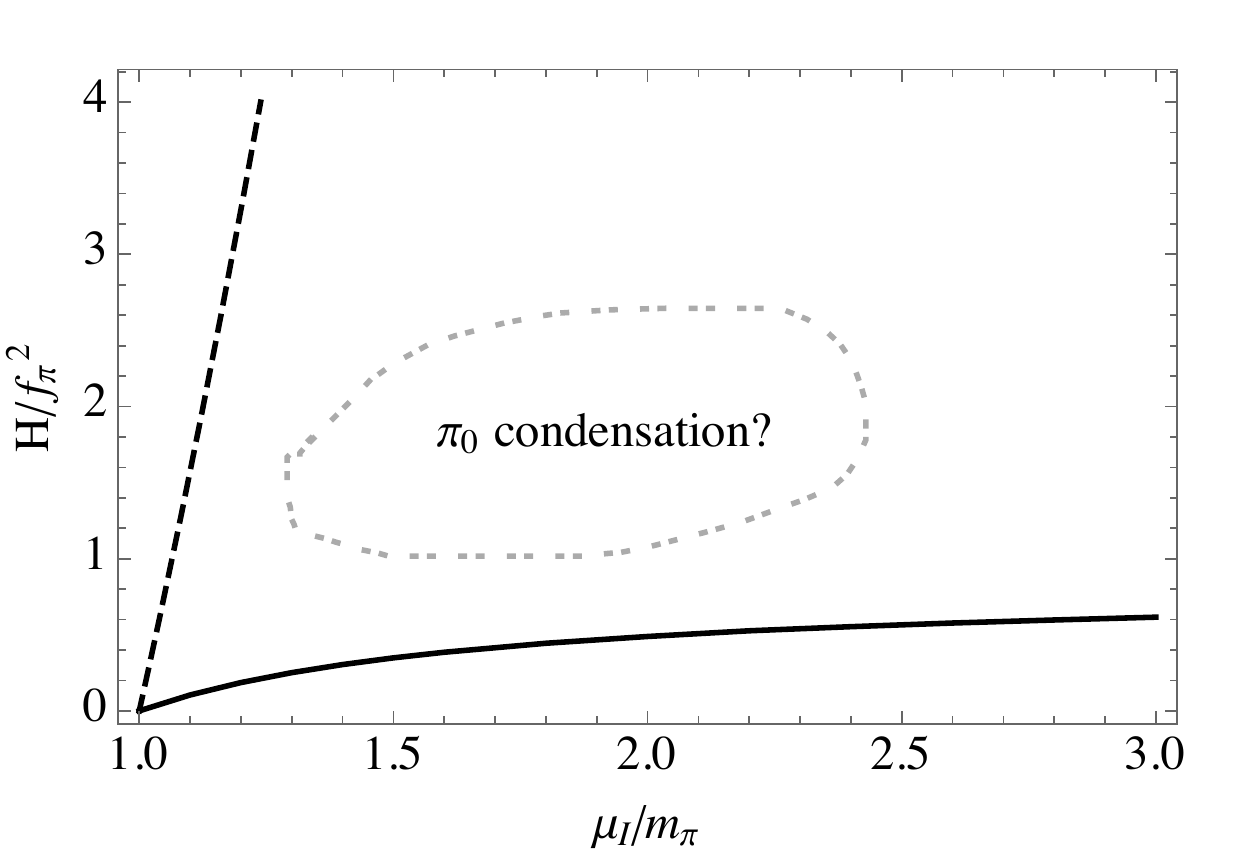}
\caption{The plot shows a possible phase diagram for finite isospin chiral perturbation theory with a region of neutral pion condensation, which is ruled out near the two critical magnetic fields.}
\label{conjecture}
\end{figure}

In this work, we have studied finite isospin $\chi PT$ in an external magnetic field (at the classical level) similar to studies of the Abelian Higgs Model. $\chi$PT is richer than the Higgs model due to the presence of derivative interactions, which emerge as a consequence of the symmetry breaking patterns of QCD and the fact that pions are Goldstone modes. There are obviously qualitative similarities both in the structure of the condensation energy and the vortex lattice structure. However, there is the added complication of the fact that there is an additional (electromagnetically) neutral degree of freedom that couples with the charged pions through strong interactions. We found that these neutral pions do not condense in the vortex lattice. Previously, we established this was the case in a single vortex~\cite{Adhikari:2015wva}. However, this does not mean that there aren't more exotic possibilities: one such example present in Fig.~\ref{conjecture}. It consists of an ``island" with neutral pions condensing in the vortex lattice away from both critical fields. 

Additionally, it is worth noting in light of the spectrum (Landau levels) of spin-0 scalar fields, in a uniform, external magnetic field that the effective mass (gap at zero momentum and zero isospin) of a charged pion becomes
\begin{equation}
\label{meff}
m^{\rm eff}_{\pi}(B_{\rm ext})=\sqrt{m_{\pi}^{2}+eB_{\rm ext}}\ .
\end{equation}
As such one might expect the transition to a diamagnetic phase to occur for $\mu_{I}\ge m^{\rm eff}_{\pi}$. This work including and our previous work in Ref.~\cite{Adhikari:2015wva} is consistent with first turning on the isospin chemical potential starting at zero until $\mu_{I}> m_{\pi}$ such that the system is in the pion condensed, superfluid state first and only subsequently turning on the external magnetic field. Since the pion superfluid is charged it behaves as a superconductor when the external field is turned on and prevents penetration by the external magnetic field.
 
In summary, we analyzed vortex lattice solutions using only the classical Lagrangian and have so far ignored the role of temperature~\cite{Haber:2017kth}, quantum fluctuations and quasi-momentum excitations of the hexagonal vortex lattice. These excitations~\cite{rosenstein2010ginzburg} are known to play an important role in determining the stability of the lattice structure and the resulting phonon spectrum. These issues will be investigated in future work.

\section*{Acknowledgements}
P.A. would like to acknowledge the hospitality of the University of Minnesota where this work was completed. P.A. would also like to acknowledge Michael Jarret for discussions on the bound state of 2d periodic potentials and Amy Kolan for discussions on the method of successive approximations in non-linear classical oscillators.
\section*{References}

\begin{thebibliography}{100}
\expandafter\ifx\csname url\endcsname\relax
  \def\url#1{\texttt{#1}}\fi
\expandafter\ifx\csname urlprefix\endcsname\relax\def\urlprefix{URL }\fi
\expandafter\ifx\csname href\endcsname\relax
  \def\href#1#2{#2} \def\path#1{#1}\fi

\bibitem{Harding:2006qn}
A.~K. Harding, D.~Lai, {Physics of Strongly Magnetized Neutron Stars}, Rept.
  Prog. Phys. 69 (2006) 2631.
\newblock \href {http://arxiv.org/abs/astro-ph/0606674}
  {\path{arXiv:astro-ph/0606674}}, \href
  {http://dx.doi.org/10.1088/0034-4885/69/9/R03}
  {\path{doi:10.1088/0034-4885/69/9/R03}}.

\bibitem{Deng:2012pc}
W.-T. Deng, X.-G. Huang, {Event-by-event generation of electromagnetic fields
  in heavy-ion collisions}, Phys. Rev. C85 (2012) 044907.
\newblock \href {http://arxiv.org/abs/1201.5108} {\path{arXiv:1201.5108}},
  \href {http://dx.doi.org/10.1103/PhysRevC.85.044907}
  {\path{doi:10.1103/PhysRevC.85.044907}}.

\bibitem{Miransky:2002rp}
V.~A. Miransky, I.~A. Shovkovy, {Magnetic catalysis and anisotropic confinement
  in QCD}, Phys. Rev. D66 (2002) 045006.
\newblock \href {http://arxiv.org/abs/hep-ph/0205348}
  {\path{arXiv:hep-ph/0205348}}, \href
  {http://dx.doi.org/10.1103/PhysRevD.66.045006}
  {\path{doi:10.1103/PhysRevD.66.045006}}.

\bibitem{Cohen:2007bt}
T.~D. Cohen, D.~A. McGady, E.~S. Werbos, {The Chiral condensate in a constant
  electromagnetic field}, Phys. Rev. C76 (2007) 055201.
\newblock \href {http://arxiv.org/abs/0706.3208} {\path{arXiv:0706.3208}},
  \href {http://dx.doi.org/10.1103/PhysRevC.76.055201}
  {\path{doi:10.1103/PhysRevC.76.055201}}.

\bibitem{Werbos:2007ym}
E.~S. Werbos, {The Chiral condensate in a constant electromagnetic field at
  O($p^6$)}, Phys. Rev. C77 (2008) 065202.
\newblock \href {http://arxiv.org/abs/0711.2635} {\path{arXiv:0711.2635}},
  \href {http://dx.doi.org/10.1103/PhysRevC.77.065202}
  {\path{doi:10.1103/PhysRevC.77.065202}}.

\bibitem{Bali:2012zg}
G.~S. Bali, F.~Bruckmann, G.~Endr{\H{o}}di, Godi, Z.~Fodor, S.~D. Katz,
  A.~Schafer, {QCD quark condensate in external magnetic fields}, Phys. Rev.
  D86 (2012) 071502.
\newblock \href {http://arxiv.org/abs/1206.4205} {\path{arXiv:1206.4205}},
  \href {http://dx.doi.org/10.1103/PhysRevD.86.071502}
  {\path{doi:10.1103/PhysRevD.86.071502}}.

\bibitem{Chernodub:2012bj}
M.~N. Chernodub, J.~Van~Doorsselaere, H.~Verschelde, {Spontaneous
  electromagnetic superconductivity and superfluidity of QCDxQED vacuum in
  strong magnetic field}, PoS QNP2012 (2012) 109.
\newblock \href {http://arxiv.org/abs/1206.2845} {\path{arXiv:1206.2845}}.

\bibitem{Chernodub:2012bq}
M.~N. Chernodub, J.~Van~Doorsselaere, H.~Verschelde, {Spontaneous
  electromagnetic superconductivity of vacuum induced by a strong magnetic
  field: QCD and electroweak theory}, AIP Conf. Proc. 1492 (2012) 281--287.
\newblock \href {http://arxiv.org/abs/1208.6118} {\path{arXiv:1208.6118}},
  \href {http://dx.doi.org/10.1063/1.4763532} {\path{doi:10.1063/1.4763532}}.

\bibitem{Son:2000by}
D.~T. Son, M.~A. Stephanov, {QCD at finite isospin density: From pion to quark
  - anti-quark condensation}, Phys. Atom. Nucl. 64 (2001) 834--842, [Yad.
  Fiz.64,899(2001)].
\newblock \href {http://arxiv.org/abs/hep-ph/0011365}
  {\path{arXiv:hep-ph/0011365}}, \href {http://dx.doi.org/10.1134/1.1378872}
  {\path{doi:10.1134/1.1378872}}.

\bibitem{Son:2000xc}
D.~T. Son, M.~A. Stephanov, {QCD at finite isospin density}, Phys. Rev. Lett.
  86 (2001) 592--595.
\newblock \href {http://arxiv.org/abs/hep-ph/0005225}
  {\path{arXiv:hep-ph/0005225}}, \href
  {http://dx.doi.org/10.1103/PhysRevLett.86.592}
  {\path{doi:10.1103/PhysRevLett.86.592}}.

\bibitem{Kogut:2000ek}
J.~B. Kogut, M.~A. Stephanov, D.~Toublan, J.~J.~M. Verbaarschot, A.~Zhitnitsky,
  {QCD - like theories at finite baryon density}, Nucl. Phys. B582 (2000)
  477--513.
\newblock \href {http://arxiv.org/abs/hep-ph/0001171}
  {\path{arXiv:hep-ph/0001171}}, \href
  {http://dx.doi.org/10.1016/S0550-3213(00)00242-X}
  {\path{doi:10.1016/S0550-3213(00)00242-X}}.

\bibitem{Splittorff:2000mm}
K.~Splittorff, D.~T. Son, M.~A. Stephanov, {QCD - like theories at finite
  baryon and isospin density}, Phys. Rev. D64 (2001) 016003.
\newblock \href {http://arxiv.org/abs/hep-ph/0012274}
  {\path{arXiv:hep-ph/0012274}}, \href
  {http://dx.doi.org/10.1103/PhysRevD.64.016003}
  {\path{doi:10.1103/PhysRevD.64.016003}}.

\bibitem{Adhikari:2018kzh}
P.~Adhikari, S.~B. Belaznay, M.~Mannarelli, {Finite Density Two Color Chiral
  Perturbation Theory Revisited}, Eur. Phys. J. C78~(6) (2018) 441.
\newblock \href {http://arxiv.org/abs/1803.00490} {\path{arXiv:1803.00490}},
  \href {http://dx.doi.org/10.1140/epjc/s10052-018-5934-6}
  {\path{doi:10.1140/epjc/s10052-018-5934-6}}.

\bibitem{Cohen:2015soa}
T.~D. Cohen, S.~Sen, {Deconfinement Transition at High Isospin Chemical
  Potential and Low Temperature}, Nucl. Phys. A942 (2015) 39--53.
\newblock \href {http://arxiv.org/abs/1503.00006} {\path{arXiv:1503.00006}},
  \href {http://dx.doi.org/10.1016/j.nuclphysa.2015.07.018}
  {\path{doi:10.1016/j.nuclphysa.2015.07.018}}.

\bibitem{Alford:1997zt}
M.~G. Alford, K.~Rajagopal, F.~Wilczek, {QCD at finite baryon density: Nucleon
  droplets and color superconductivity}, Phys. Lett. B422 (1998) 247--256.
\newblock \href {http://arxiv.org/abs/hep-ph/9711395}
  {\path{arXiv:hep-ph/9711395}}, \href
  {http://dx.doi.org/10.1016/S0370-2693(98)00051-3}
  {\path{doi:10.1016/S0370-2693(98)00051-3}}.

\bibitem{Alford:1998mk}
M.~G. Alford, K.~Rajagopal, F.~Wilczek, {Color flavor locking and chiral
  symmetry breaking in high density QCD}, Nucl. Phys. B537 (1999) 443--458.
\newblock \href {http://arxiv.org/abs/hep-ph/9804403}
  {\path{arXiv:hep-ph/9804403}}, \href
  {http://dx.doi.org/10.1016/S0550-3213(98)00668-3}
  {\path{doi:10.1016/S0550-3213(98)00668-3}}.

\bibitem{Adhikari:2018cea}
P.~Adhikari, J.~O. Andersen, P.~Kneschke, {Pion condensation and phase diagram
  in the Polyakov-loop quark-meson model}\href
  {http://arxiv.org/abs/1805.08599} {\path{arXiv:1805.08599}}.

\bibitem{Endrodi:2014lja}
G.~Endr{\H{o}}di, {Magnetic structure of isospin-asymmetric QCD matter in
  neutron stars}, Phys. Rev. D90~(9) (2014) 094501.
\newblock \href {http://arxiv.org/abs/1407.1216} {\path{arXiv:1407.1216}},
  \href {http://dx.doi.org/10.1103/PhysRevD.90.094501}
  {\path{doi:10.1103/PhysRevD.90.094501}}.

\bibitem{Adhikari:2015wva}
P.~Adhikari, T.~D. Cohen, J.~Sakowitz, {Finite Isospin Chiral Perturbation
  Theory in a Magnetic Field}, Phys. Rev. C91~(4) (2015) 045202.
\newblock \href {http://arxiv.org/abs/1501.02737} {\path{arXiv:1501.02737}},
  \href {http://dx.doi.org/10.1103/PhysRevC.91.045202}
  {\path{doi:10.1103/PhysRevC.91.045202}}.

\bibitem{Adhikari:2013dfa}
P.~Adhikari, T.~D. Cohen, I.~Datta, {Density of Saturated Nuclear Matter at
  Large $N_{c}$ and Heavy Quark Mass Limits}, Phys. Rev. C89~(6) (2014) 065201.
\newblock \href {http://arxiv.org/abs/1312.3339} {\path{arXiv:1312.3339}},
  \href {http://dx.doi.org/10.1103/PhysRevC.89.065201}
  {\path{doi:10.1103/PhysRevC.89.065201}}.

\bibitem{tinkham2004introduction}
M.~Tinkham, Introduction to Superconductivity, Vol.~1, McGraw-Hill, New York,
  1996.

\bibitem{Babichev:2007tn}
E.~Babichev, {Gauge k-vortices}, Phys. Rev. D77 (2008) 065021.
\newblock \href {http://arxiv.org/abs/0711.0376} {\path{arXiv:0711.0376}},
  \href {http://dx.doi.org/10.1103/PhysRevD.77.065021}
  {\path{doi:10.1103/PhysRevD.77.065021}}.

\bibitem{Adhikari:2017oxb}
P.~Adhikari, J.~Choi, {Vortex solutions in the Abelian Higgs Model with a
  neutral scalar}, Acta Phys. Polon. B48 (2017) 145.
\newblock \href {http://arxiv.org/abs/1703.00961} {\path{arXiv:1703.00961}},
  \href {http://dx.doi.org/10.5506/APhysPolB.48.145}
  {\path{doi:10.5506/APhysPolB.48.145}}.

\bibitem{Chernodub:2011gs}
M.~N. Chernodub, J.~Van~Doorsselaere, H.~Verschelde, {Electromagnetically
  superconducting phase of vacuum in strong magnetic field: structure of
  superconductor and superfluid vortex lattices in the ground state}, Phys.
  Rev. D85 (2012) 045002.
\newblock \href {http://arxiv.org/abs/1111.4401} {\path{arXiv:1111.4401}},
  \href {http://dx.doi.org/10.1103/PhysRevD.85.045002}
  {\path{doi:10.1103/PhysRevD.85.045002}}.

\bibitem{Abrikosov:1956sx}
A.~A. Abrikosov, {On the Magnetic properties of superconductors of the second
  group}, Sov. Phys. JETP 5 (1957) 1174--1182, [Zh. Eksp. Teor.
  Fiz.32,1442(1957)].

\bibitem{Adhikari:2018tvz}
P.~Adhikari, J.~Choi, {Magnetic Vortices in the Abelian Higgs Model with
  Derivative Interactions}\href {http://arxiv.org/abs/1810.00917}
  {\path{arXiv:1810.00917}}.

\bibitem{abrikosov1988fundamentals}
A.~A. Abrikosov, Fundamentals of the Theory of Metals, North Holland,
  Amsterdam, 1988.

\bibitem{kleiner1964bulk}
W.~Kleiner, L.~Roth, S.~Autler, Bulk solution of ginzburg-landau equations for
  type ii superconductors: upper critical field region, Phys. Rev. 133~(5A)
  (1964) A1226.

\bibitem{Zak:1964zz}
J.~Zak, {Magnetic Translation Group}, Phys. Rev. 134 (1964) A1602--A1606.
\newblock \href {http://dx.doi.org/10.1103/PhysRev.134.A1602}
  {\path{doi:10.1103/PhysRev.134.A1602}}.

\bibitem{avron1978separation}
J.~Avron, I.~Herbst, B.~Simon, Separation of center of mass in homogeneous
  magnetic fields, Annals of Physics 114~(1-2) (1978) 431--451.

\bibitem{rosenstein2010ginzburg}
B.~Rosenstein, D.~Li, Ginzburg-landau theory of type ii superconductors in
  magnetic field, Rev. Mod. Phys. 82~(1) (2010) 109.

\bibitem{coleman1988aspects}
S.~Coleman, Aspects of Symmetry: Selected Erice Lectures, Cambridge University
  Press, Cambridge, 1988.

\bibitem{Haber:2017kth}
A.~Haber, A.~Schmitt, {Critical magnetic fields in a superconductor coupled to
  a superfluid}, Phys. Rev. D95~(11) (2017) 116016.
\newblock \href {http://arxiv.org/abs/1704.01575} {\path{arXiv:1704.01575}},
  \href {http://dx.doi.org/10.1103/PhysRevD.95.116016}
  {\path{doi:10.1103/PhysRevD.95.116016}}.

\end{thebibliography}

\end{document}